\documentclass[11pt]{article}
\usepackage{float}
\usepackage[symbol]{footmisc}
\usepackage[superscript,sort]{cite}
\usepackage{array}
\usepackage[normalem]{ulem} 
\usepackage{graphicx}
\usepackage[colorlinks=true,linkcolor=black,citecolor=black,urlcolor=black,bookmarks=true,breaklinks=true]{hyperref}

\usepackage{mathtools}
\usepackage{amsthm,amscd,amsxtra,amsfonts,amsmath,amssymb,multirow}
\usepackage{wrapfig}
\usepackage[footnotesize]{caption}
\usepackage{subcaption}
\usepackage[tiny,compact]{titlesec}
\usepackage{booktabs}
\usepackage{siunitx}
\usepackage{threeparttable} 
\usepackage{tabularx}
\usepackage{helvet}

\usepackage[dvipsnames]{xcolor}
\usepackage{colortbl}
\usepackage{tikz}
\usetikzlibrary{shapes,arrows}
\usepackage[T1]{fontenc}
\usepackage[linesnumbered,ruled]{algorithm2e} 
 
\titlespacing*{\section}{0pt}{*0}{*0}
\titlespacing*{\subsection}{0pt}{*0}{*0}
\titlespacing*{\subsubsection}{0pt}{*0}{*0}
\titlespacing{\paragraph}{0pt}{*0}{*1}

\usepackage{mathtools}
\begin{document}
\pagenumbering{roman}

\clearpage \pagebreak \setcounter{page}{1}
\renewcommand{\thepage}{{\arabic{page}}}

\title{  MathDL: Mathematical deep learning for D3R Grand Challenge 4
}

\author{Duc Duy Nguyen$^{1}$,  Kaifu Gao$^{1}$, Menglun Wang$^{1}$ and  Guo-Wei Wei$^{1,2,3,}$\footnote{
		Corresponding to Guo-Wei Wei.		Email: weig@msu.edu}\\
	$^1$ Department of Mathematics,
	Michigan State University, MI 48824, USA.\\
	$^2$ Department of Biochemistry and Molecular Biology,
	Michigan State University, MI 48824, USA. \\
$^3$ Department of Electrical and Computer Engineering,
	Michigan State University, MI 48824, USA. \\
}

\date{\today}
\maketitle

\begin{abstract}
We present the performances of our mathematical deep learning (MathDL) models for D3R Grand Challenge 4 (GC4).
This challenge involves pose prediction, affinity ranking, and free energy estimation for beta secretase 1 (BACE) as well as affinity ranking and free energy estimation for Cathepsin S (CatS). We have developed advanced mathematics, namely differential geometry, algebraic graph, and/or algebraic topology,  to accurately and efficiently encode  high dimensional physical/chemical interactions into {  scalable low-dimensional rotational and translational invariant representations.} These representations are integrated with deep learning models, such as generative adversarial networks (GAN) and convolutional neural networks (CNN) for pose prediction and energy evaluation, respectively. Overall, our MathDL models achieved the top place in pose prediction for BACE ligands in Stage 1a. Moreover, our submissions obtained the highest Spearman correlation coefficient on the affinity ranking of 460 CatS compounds, and the smallest centered root mean square error on the free energy set of 39 CatS molecules. { It is worthy to mention that our method for docking pose predictions has significantly improved from our previous ones.}

\end{abstract}
 
\maketitle



\newpage

\section{Introduction}
The Drug Design Data Resource (D3R) offers blind communitywide challenges of ligand pose and binding affinity ranking predictions \cite{gathiaka2016d3r,gaieb2018d3r,gaieb2019d3r}. Benchmarks in D3R contests contain high quality structures and reliable binding energies supplied by experimental groups before the publication. These challenges provide computer-aided drug design (CADD) community a great opportunity to validate, calibrate, and develop drug virtual screening (VS) models. The latest
D3R Grand Challenge 4 (GC4), took place from September 4th 2018 to December 4th, 2018. GC4 presented two different protein targets, Cathepsin S (CatS) and beta secretase 1 (BACE), which were generously supplied by Janssen Pharmaceuticals and Novartis, respectively. There were two stages in GC4. The first one has two subchallenges, namely Stage 1a and Stage 1b. In Stage 1a, participants were asked to predict the pose, rank the affinity, and estimate the free energy of BACE ligands. Following Stage 1a, Stage 1b revealed the receptor structures and participants were asked again to predict the crystallographic poses of  20 BACE ligands. There was no affinity calculation in this stage { 1b}. The second part of GC4 was called Stage 2 which contained the affinity rankings and free energy challenges for both BACE and CatS compounds. In this last stage, participants were able to take advantage of experimental structures of BACE complexes released right after stage 1b.

A successful VS model requires a reliable ligand conformation generation and highly accurate scoring function to predict binding affinities. There are several state-of-the-art software packages to take care of the first component of VS, for example, Autodock Vina \cite{Trott:2010AutoDock}, GOLD \cite{G-Score}, GLIDE \cite{Friesner:2004Schrodinger}, ICM \cite{abagyan1994icm}, etc. Unfortunately, one may fail dramatically to achieve decent poses if blindly using these software programs. The pose prediction results in Grand Challenge 3 (GC3) clearly { demonstrated this issue} \cite{gaieb2019d3r}. The second component of VS relates to the development of scoring function (SF) for binding affinity predictions. Basically, one can classify { SF methods} into four different types, namely force-field-based SF, knowledge-based SF, empirical-based SF, and machine learning-based SF \cite{LiuJie:2014}. The force-field-based SFs commonly emphasize van der Walls (vdW) interactions, electrostatic energy, hydrogen bonding descriptions, solvation effects, and so on. The well-known SFs for this category are COMBINE \cite{Ortiz:1995}, MedusaScore \cite{Yin:2008}, to name only a few. { Typical examples of} knowledge-based SFs are \cite{PMFScore:1999}, DrugScore\cite{DrugScore:2005}, KECSA\cite{Merz:2015Solvation}, and IT-Score \cite{ITScore:2006}, which utilize {  protein-ligand pairwise statistical potentials} in an additive manner to predict  binding affinities. One can regard the empirical-based SFs as simple machine learning-based SFs since these SFs  employ {   linear regression schemes to construct  predictive models  using various} physical features, for instance vdW interactions, Lennard-Jones potentials, hydrogen bonds, electrostatics, solvation, and torsion information, etc. PLP\cite{Verkhivker:1995PLP}, ChemScore \cite{Eldridge:1997}, and X-Score \cite{XScore:2002} are the well-known representatives { in this category. The last type of binding affinity SFs} is machine learning-based approaches which have recently arise as the most advanced technique in CADD. One of the pioneer work on this SF category is RF-Score \cite{Pedro:2010Binding} based on the Random Forest (RF) algorithm \cite{Breiman:2001} and their features as the numbers of atom pairwise contacts. Thanks to the nonlinear representation of the sophisticated machine learning frameworks, machine learning-based SFs can characterize the non-additive contributions from functional group interactions in the binding affinity calculations \cite{li:2014,DDNguyen:2017d,ZXCang:2017b,ZXCang:2017c,ZXCang:2018a,nguyen2019dg,nguyen2019agl}.

The availability of massive biological datasets, along with the accessibility to high-performance computing cluster (HPCC), has made machine learning-based models an emerging technology in biomolecular data analysis and prediction. However, the accuracy of machine learning-based SFs highly depends on whether their features are able to capture the physical and chemical information in protein-ligand interactions. Moreover, the direct use of three dimensional (3D) biomolecular structures in the deep learning network is immensely expensive. This hindrance mainly causes by the hefty number of degrees of freedom in the 3D macromolecular representations and the number of atoms varying among different structures. Therefore, there is a pressing need to develop innovative { representations of protein-ligand complexes for } machine learning  methods.

Mathematical deep learning (MathDL) encompasses a family of  { scalable  low-dimensional rotational and translational invariant  mathematical representations integrated with advanced machine learning,  including} deep learning algorithms \cite{nguyen2019mathematical}. Its hypothesis is that the intrinsic physics of macromolecular interactions lie in  low-dimensional manifolds.  Based on such hypothesis, we have developed a number of mathematical tools originated from geometry, topology, graph theory,  combinatorics, and analysis to simplify macromolecular complexity and reduce their dimensionality. For example, differential geometry provides a high-level abstraction of macromolecular complexes \cite{Wei:2009}. In molecular biophysics, differential geometry-based framework has shown its efficiency in modeling solvation-free energies \cite{ZhanChen:2012,BaoWang:2015cOptParm} and ion channel transport \cite{DuanChen:2011d,DuanChen:2012b,Wei:2012,Wei:2013,DuanChen:2013}. However, in those applications, differential geometry information is largely restricted to the separation of solvent and solute domains in facilitating the Poisson-Boltzmann model or the Poisson-Nernst-Planck model. In geometric modeling, differential geometry has been utilized for the qualitative analysis of biomolecule properties \cite{XFeng:2012a,KLXia:2014a}. Also, potential protein-ligand binding sites can be recognized via concave and convex regions of molecular surfaces indicated by  minimum and/or maximum curvatures \cite{KLXia:2014a,mu2017geometric}. Most recently, the roles of different kinds of curvature in solvation free energy models have been investigated \cite{DDNguyen:2016c}. However, the efficiency of the aforementioned  differential geometry models is limited due to neglecting of atomic level information. Element interactive manifolds (EIM) were proposed to address this problem in differential geometry-based geometric learning (DG-GL) \cite{nguyen2019dg}.  These EIMs successfully encode the pivotal physical, chemical, and biological information stored in high-dimensional data into low-dimensional manifolds, rendering  a powerful approach for  predicting  solvation free energy, drug toxicity, and protein-ligand binding affinity \cite{nguyen2019dg}.

Another low-dimensional mathematical approach is the topological representation of biomolecular structures. In topological data analysis, one can capture the connectivity of  macromolecules or molecular components. Topological invariants, such as independent components, rings, cavities, and higher dimension faces in terms of Betti numbers help to characterize the conformation change upon the protein-ligand binding process, the folding and unfolding of proteins, and the opening or closing of ion channels \cite{Kaczynski:2004}. The traditional topological descriptors, unfortunately, cannot discriminate the geometric difference among various macromolecular structures. Persistent homology (PH), a new branch of algebraic topology, utilizes a filtration parameter to generate a family of  topological spaces and associated invariants, which contain richer  geometric information \cite{Edelsbrunner01topologicalpersistence, Zomorodian:2005}.  PH has been applied to computational biology \cite{Kasson:2007,dabaghian2012topological, Gameiro:2014}. However, these applications were mostly limited to qualitative analysis. Recently, we have devised PH  for the quantitative analysis of protein folding energy,  protein flexibility \cite{KLXia:2014c}, ill-posed inverse problems of cryo-EM structures \cite{KLXia:2015b}, predictive models of curvature energies of fullerene isomers \cite{KLXia:2015a,BaoWang:2016a}, and protein pocket detection \cite{liu2017eses}.  In 2015, we introduced one of the first combinations of PH descriptors and machine learning algorithms \cite{ZXCang:2015}. Since then, the integration of PH and machine learning has become a very popular approach in topological data analysis. Nonetheless, this approach is not good enough for biomolecular systems. It turns out that PH neglects chemical and biological information in its topological simplification of geometric complexity. Element-specific PH was introduced to retain chemical and biological information  \cite{ZXCang:2017b}. The integration of  element-specific PH and machine learning algorithms has found great success in the predictions of protein folding free energy changes upon mutation \cite{ZXCang:2017a}, binding affinity \cite{ZXCang:2017b,ZXCang:2017c,ZXCang:2018a},    drug toxicity  \cite{KDWu:2018a}, partition coefficient, and aqueous solubility \cite{KDWu:2018b}. It has been employed for the classification of active ligands and decoys \cite{ZXCang:2018a}. All of these new topological models outperformed other state-of-the-art methods on various common benchmarks.

Similarly to topology, graph theory also accentuates the connectivity between vertices to define graph edges. {  There are two major types of graphs: geometric graphs and algebraic graphs. Geometric graphs concern  the pairwise connectivity between  graph nodes and represent it in terms of } ``topological index'' \cite{hosoya1971topological,hansen1988chemical}, graph centrality \cite{newman2010networks,bavelas1950communication,dekker2005conceptual}, and contact map \cite{Bahar:1997,LWYang:2008}. The algebraic graph theory expresses the connectivity via { eigenvalues, particularly,} the second-smallest eigenvalue of the Laplacian matrix, known as Fiedler value, which is often used to analyze the stability of dynamical systems \cite{Wei:2002e}. Graph theory has been widely used in many interdisciplinary studies. In biophysics, it is employed to model protein flexibility and long-time dynamics in normal mode analysis (NMA) \cite{Go:1983,Tasumi:1982,Brooks:1983,Levitt:1985} and elastic network model (ENM)\cite{Bahar:1997,Flory:1976,Bahar:1998,Atilgan:2001,Hinsen:1998,Tama:2001,QCui:2010}. Since graph theory offers a nature representation of molecular structure, it is a common approach for analyzing chemical  datasets \cite{balaban1976chemical,trinajstic1983chemical, schultz1989topological, foulds2012graph, hansen1988chemical, ozkanlar2014chemnetworks} and biomolecular datasets \cite{Bahar:1997,di2015protein, canutescu2003graph, ryslik2014graph,Jacobs:2001, vishveshwara2002protein,wu2017moleculenet}. Although there was much effort in constructing various graph representations in the past,  graph based quantitative models are often less accurate than other competitive models in  {   the analysis and prediction of biomolecular properties from massive and diverse datasets}. Indeed, in the protein stability changes upon mutation analysis, the other models \cite{LQuan:2016,ZXCang:2017a, ZXCang:2017c} are more accurate than the graph-based approach \cite{Pires:2014}. In addition, the graph theory based Gaussian network model (GNM) is not competitive in protein  B-factor predcitions  \cite{JKPark:2013}. One of the main reasons is that there is no systematic representation of interactions  among different chemical element types in a molecular structure. Additionally, many graph approaches do not describe  non-covalent interactions.  To overcome these limitations, we have proposed novel  multiscale weighted colored subgraphs in  both  geometric graph and algebraic graph schemes to achieve the state-of-the-art performances in the predictions of protein B-factor \cite{DBramer:2018a}, protein-ligand binding affinity \cite{DDNguyen:2017d,nguyen2019agl}, docking \cite{nguyen2019agl}, and virtual screening \cite{nguyen2019agl}.

Our  MathDL models using graph theory and algebraic topology were employed in the D3R Grand Challenges since GC2 and has obtained many encouraging results. Specifically, our prediction of the free energy set in Stage 2 was ranked the best in GC2 in our first participation of D3R competitions \cite{nguyen2019mathematical}. In our second participation,  i.e. GC3, our submissions achieved the top places in 10 out of 26 official contests \cite{nguyen2019mathematical}.  These achievements have confirmed the predictive power and efficiency of our MathDL models in drug design and discovery. However, there were still some shortcomings existing in our previous approaches mostly concerning the pose generation performance and ability to rank affinities of compounds with diverse chemical structures.

In the current D3R challenge, i.e. GC4, we have brought in two new technological aspects in our approach. First, we have further developed powerful differential geometry and algebraic graph-based MathDL models to assist our algebraic topology based methods. Additionally, we  have extended our MathDL approach with more advanced  deep learning architectures like generative adversarial networks (GAN) \cite{goodfellow2014generative}.  We have achieved very promising results with top places in pose prediction, affinity ranking and free energy estimation. The rest of this paper is devoted to more detailed discussions of our methodologies and their performances in D3R GC4.

 \section{Methods}

We describe the {   mathematical methods }  underpinning our MathDL models in this sections.

\subsection{Differential geometry representation}\label{sec:DG}
\subsubsection{Multiscale discrete-to-continuum mapping} \label{Density}

Given a molecule having $N$ atoms. Denote ${\bf r}_i$ and $q_j$, $i=1\cdots N$, respectively, an atomic coordinate and a partial charge of the $j$th atom. A discrete-to-continuum mapping \cite{KLXia:2013d,Opron:2014,DDNguyen:2016b} represents the unnormalized molecular density at an arbitrary point ${\bf r} \in \mathbb{R}^3$ as follows
 \begin{align}\label{fri_surface}
    \rho(\mathbf{r}, \{ \eta_k\}, \{w_k\})=\sum_{\substack{j=1}}^{N} w_j  \Phi\left(\|\mathbf{r}-\mathbf{r}_j\|;\eta_{j}\right),
\end{align}
where $\|\mathbf{r} -\mathbf{r}_j\|$ is  the Euclidean distance of the point ${\bf r}$ and the $j$th atom in a given molecule. If all $w_j$ are  set to 1, $\rho(\mathbf{r}, \{ \eta_k\}, \{w_k\})$ indicates a molecular density, whereas $\rho(\mathbf{r}, \{ \eta_k\}, \{w_k\})$ serves as molecular charge density with $w_j=q_j$ for all $j$. { In the present work, we utilize Autodock Tools (\url{http://autodock.scripps.edu/resources/adt/index_html}) to assign the Gasteiger charges for small molecules and macromolecules.} Additionally, $ \eta_{j}$ are characteristic distances and $\Phi$ is a monotonically decreasing kernel featuring the similarity between two 3D data points.  {  To ensure the existence of the geometric representations such as curvatures}, {  $\Phi$ is chosen to be monotonically decreasing $C^2$ function satisfying the following conditions}
\begin{align}
\Phi \left(\|\mathbf{r}- \mathbf{r}_j\|;\eta_{j}\|\right)&=1, \quad{\rm as} \quad  \|\mathbf{r} -\mathbf{r}_j\| \rightarrow 0, \label{eq:admiss-1}\\
\Phi \left(\|\mathbf{r} - \mathbf{r}_j\|;\eta_{j}\|\right)&=0, \quad {\rm as} \quad  \|\mathbf{r} -\mathbf{r}_j\| \rightarrow \infty. \label{eq:admiss-2}
\end{align}
It is noted that {   radial basis functions meet admissibility conditions} (\ref{eq:admiss-1}) and (\ref{eq:admiss-2}). Commonly used correlation kernels are
generalized exponential functions
\begin{align}\label{exponential}
\Phi\left(\|\mathbf{r} -\mathbf{r}_j\|;\eta_{ j}\|\right)=e^{-\left(\|\mathbf{r} -\mathbf{r}_j\|/\eta_{ j}\right)^\kappa}, \quad \kappa>0;
\end{align}
and generalized Lorentz functions
\begin{align}\label{Lorentz1}
\Phi\left(\|\mathbf{r} -\mathbf{r}_j\|;\eta_{ j}\right)=\frac{1}{1+\left(\|\mathbf{r} -\mathbf{r}_j\|/\eta_{ j}\right)^\nu},\quad \nu>0.
\end{align}
{ 
Moreover, one can use correlation kernels to model the electrostatic interaction between two charged articles as the following
\begin{align}\label{electrostatic}
\Phi(\|\mathbf{r}_i -\mathbf{r}_j\|, q_i, q_j; c) = \frac{1}{1+e^{-cq_iq_j/\|\mathbf{r}_i -\mathbf{r}_j\|}},
\end{align}
where, $q_i$ and $q_j$ are the partial charges of two atoms, and $c$ is a nonzero tunable parameter. All the $\Phi$s discussed in the current work were determined by one of Eqs. (\ref{exponential}) - (\ref{electrostatic}). Here, $\Phi$ takes 3D coordinates and kernel parameters as the input variables and maps them to a real number: $\mathbb{ R}^3\rightarrow \mathbb{ R}$. Therefore, $\Phi$ values totally depend on atom coordinates or grid point positions and are rotationally and translationally invariant.
}

It is expected that $C^2$ delta sequences of the positive type discussed in an earlier work \cite{GWei:2000} can function well for the correlation kernel purposes. To obtain multiscale discrete-to-continuum mapping, one can employ more than one set of scale parameters.  { In the current work, the aforementioned mapping was applied to protein-ligand complexes.}

\subsubsection{Element interactive densities }  \label{sec:EID}
{ 
In order for differential geometry (DG) representations to effectively capture the crucial physical and biological information of large and diverse biomolecular datasets, we must employ DG to feature non-covalent intramolecular molecular interactions in a molecule and intermolecular interactions in molecular complexes, such as protein-protein and protein-ligand. }

Additionally, the accuracy of the DG representations  can be upgraded by element-level descriptions which result in scalable low-dimension manifold representations of high dimensional structures. For instance, to describe the pairwise interactions between protein and ligand, we consider frequently occurring element types in proteins and ligands. Particularly,  the commonly occurring element types in proteins are ${\rm C, N, O, S}$  and commonly occurring element types in ligands are ${\rm H, C, N, O, S, P, F, Cl, Br, I}$. That gives rise to 40 element pairwise groups. We do not include hydrogen in protein element types since ${\rm H}$ is usually absent from {   most datasets in the} Protein Data Bank (PDB). {  Note that during our validation process, the pairwise interactions between different atom types did not enhance the overall performance of our models (this may be due to the limited data size.). Thus, we only carried out the element-specific interactions for the sake of simplicity.}

{ 
Based on a statistical analysis, the } frequently occurring element types in the biomolecular dataset are denoted as    ${\cal C}=\{{\rm H, C, N, O, S, P, F, Cl,  \cdots }\}$. For convenience, ${\cal C}_k$ represents the $k$th element in the set ${\cal C}$. For example, ${\cal C}_5= {\rm S}$. An $i$th atom in a given molecule is associated with its coordinate ${\mathbf{r}_i}$, element type $\alpha_i$, and partial charge $q_i$. The non-covalent interactions between atoms of element type ${\cal C}_k$  and ${\cal C}_{k'}$ are assumed to be described by the correlation kernel $\Phi$

\begin{equation}\label{CollInter}
\{\Phi(||\mathbf{r}_i-\mathbf{r}_j||; \eta_{kk'})| \alpha_i  = {\cal C}_{k}, \alpha_j  = {\cal C}_{k'};  i,j = 1,2,\ldots,N;
||\mathbf{r}_i-\mathbf{r}_j||> r_i+r_j +\sigma \},
\end{equation}
 where $r_i $ and $ r_j$ are the atomic radii of $i^{th}$ and $j^{th}$ atoms, respectively and $\sigma$ is the mean value of the standard deviations of $r_i $ and $ r_j$  in the interested dataset. The covalent interactions are excluded due to the constraint $||\mathbf{r}_i-\mathbf{r}_j||> r_i+r_j +\sigma $. In addition,
 $\eta_{kk'}$ is a characteristic distance between the atoms, which depends only on their element types.

To construct the element interactive densities, we define {\it atomic-radius-parametrized} van der Waals domain of all atoms of $k$th element type as \cite{nguyen2019dg}
\begin{equation}
D_k:= \cup_{{\bf r}_i, \alpha_i = {\cal C}_k } B({\bf r}_i, r_k),
\end{equation}
in which $B({\bf r}_i,r_i)$ is a ball with a center ${\bf r}_i$ and a radius $r_i$, and $r_k$ is the atomic radius of the $k$th element type. {  Thus, $D_k$ depends on atom coordinate ${\bf r_i}$ and its atomic radius. Note that, $D_k$ does not define any vdW interactions but a domain to construct the surface density}. The element interactive density between domain $D_k$ and all atoms of $k'$th  ($k\neq k'$) element type is given by
\begin{align}\label{ESRI}
\rho_{kk'}({\bf r},  \eta_{kk'}) =\sum_{\mathclap{\substack{j \\  \alpha_j  = {\cal C}_{k'} \\
||\mathbf{r}_i-\mathbf{r}_j||> r_i+r_j +\sigma, \forall \alpha_i\in    {\cal C}_{k}
}}}
w_j \Phi(||\mathbf{r}-\mathbf{r}_j||;\eta_{kk'}),
\quad {\bf r}  \in D_k.
\end{align}

When $k'=k$, the element interactive density $\rho_{kk}$ is now induced only by van der Waals domain $D_k$. In this case, we exclude the covalent interactions based on the position of the density input. Assuming ${\bf r}  \in D^i_k$, with $D^i_k=B({\bf r}_i,r_i)$, $\alpha_i = {\cal C}_k$, the element interactive density is then formulated by

\begin{equation}\label{ESRI2}
\rho_{kk}({\bf r}, \eta_{kk}) =  \sum_{\mathclap{\substack{
j \\ \alpha_j  = {\cal C}_{k} \\ ||\mathbf{r}_i-\mathbf{r}_j||> 2r_j +\sigma}
}
}
w_j \Phi(||\mathbf{r}-\mathbf{r}_j||; \eta_{kk}).
\end{equation}

\begin{figure}[!ht]
    \centering
    \includegraphics[width=0.6\textwidth]{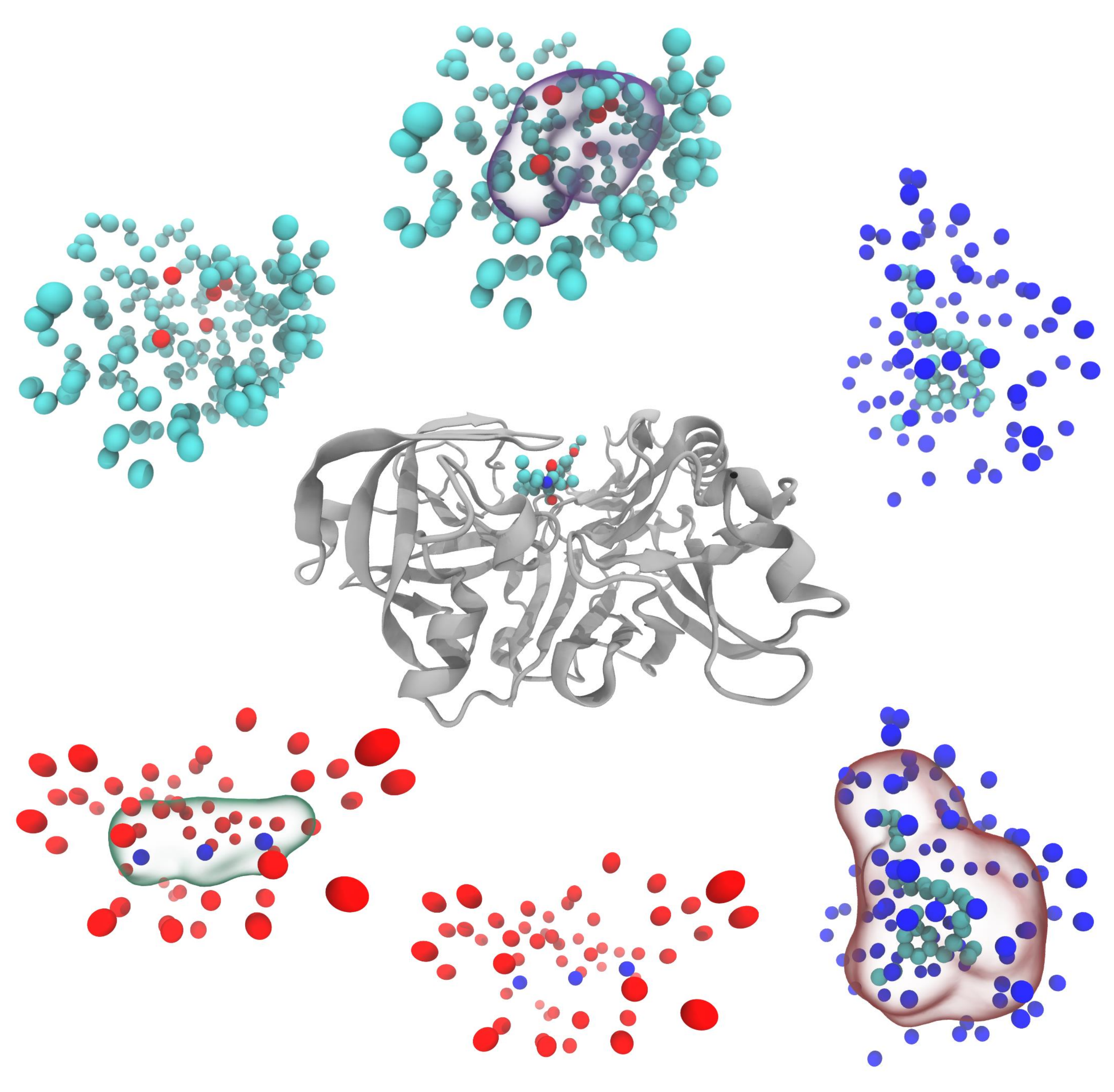}
    \caption{IIlustration of some element-specific selections and corresponding element interactive manifolds obtained at a given level set of the element interactive density. { Each sphere illustrates the atomic positions. Cyan, red, and blue colors represent carbon, oxygen, and nitrogen, respectively. The transparent surfaces are the isosurface extracted from volume data represented in Eq. (8).
    }
    }
    \label{fig:DG}
\end{figure}

{ For the sake of simplicity, we chose $w_j=1$ for all cases. Since element interactive density is obtained by the addition of correlation kernels,} it belongs to $C^2$ on the closed domain of $D_k$. We construct element interactive manifolds by restricting the set of points at a given level set of the density as shown in Fig. \ref{fig:DG} .

\subsubsection{Element interactive curvatures}

Given an element interactive density $\rho({\bf r})$, one can calculate the Gaussian curvature ($K$), the mean curvature ($H$), the minimum curvature ($\kappa_{\rm min}$), and the maximum curvature ($\kappa_{\rm max}$) for the resulting manifold as the following \cite{Soldea:2006,KLXia:2014a}:
    \begin{align}
    K=&\frac{1}{g^2}\left[
    2\rho_x \rho_y \rho_{xz}\rho_{yz} + 2\rho_x\rho_z\rho_{xy}\rho_{yz}+2\rho_y\rho_z\rho_{xy}\rho_{xz} \right. \nonumber\\
    &\left.  - 2 \rho_x \rho_z \rho_{xz} \rho_{yy} - 2 \rho_y \rho_z \rho_{xx} \rho_{yz} - 2 \rho_x \rho_y \rho_{xy} \rho_{zz} \right.\nonumber\\
    &\left. +\rho_z^2 \rho_{xx}  \rho_{yy} + \rho_x^2 \rho_{yy} \rho_{zz} + \rho_y^2 \rho_{xx} \rho_{zz}\right.\nonumber\\
    &\left. -\rho_x^2 \rho_{yz}^2 - \rho_y^2 \rho_{xz}^2 - \rho_z^2 \rho_{xy}^2 \right],
    \label{gaussian_curv}\\
    H=&\frac{1}{2g^{\frac{3}{2}}}\left[
    2 \rho_x \rho_y \rho_{xy} + 2 \rho_x \rho_z \rho_{xz} + 2 \rho_y \rho_z \rho_{yz} - (\rho_y^2 + \rho_z^2)\rho_{xx} - (\rho_x^2 + \rho_z^2)\rho_{yy} - (\rho_x^2+\rho_y^2)\rho_{zz}\right],
    \label{mean_curv}\\
    \kappa_{\rm min}=&H-\sqrt{H^2-K},\label{min_curv}\\
    \kappa_{\rm max}=&H+\sqrt{H^2-K},\label{max_curv}
    \end{align}
    where $g=\rho_x^2 + \rho_y^2 + \rho_z^2$.

To construct unified curvature quantities for various biomolecular structures, we study the element interactive curvatures (EIC) at the atomic center and formulate them as \cite{nguyen2019dg}

\begin{equation}\label{InterCurv}
   K^{\rm EI}_{kk'}(\eta_{kk'}) = \sum_i  K_{kk'}({\bf r}_i, \eta_{kk'}), \quad {\bf r}_i\in D_k; k\neq k'
\end{equation}
and
\begin{equation}\label{InterCurv2}
   K^{\rm EI}_{kk}(\eta_{kk}) =   \sum_i  K_{kk}({\bf r}_i, \eta_{kk}), \quad {\bf r}_i\in D^i_k, D^i_k \subset D_k.
\end{equation}

Eqs. \eqref{InterCurv} and \eqref{InterCurv2} are for the element interactive Gaussian curvature (EIGC), {   are applied to  protein-ligand complexes in the current work. Thus, the atomic centers in Eqs. \eqref{InterCurv} and \eqref{InterCurv2} can be either from ligand atoms or protein atoms}. In a same manner, one can define $ H^{\rm EI}_{kk'}(\eta_{kk'}),          \kappa^{\rm EI}_{kk',\rm min}(\eta_{kk'})$ and $ \kappa^{\rm EI}_{kk',\rm max}( \eta_{kk'})$ for the element interactive mean curvature, element interactive minimum curvature, and element interactive maximum curvature, respectively.

It is worth noting that, the expressions of the curvatures defined in \eqref{gaussian_curv} , \eqref{mean_curv}, \eqref{min_curv}, and \eqref{max_curv} are in the analytical forms. Thus, the EIC formulations are free from numerical error and totally preserve the reference geometric information of the molecules.

\subsection{Multiscale weighted colored geometric subgraphs }\label{sec:geometric_subgraph}
For a given molecular datasets, we denote ${\mathcal C}$ a set consisting of the most frequently appearing element types. For a  molecule of interest, we define a graph with the following vertices
\begin{equation}
{\mathcal V} = \{(\mathbf{r}_j, \alpha_j)|\mathbf{r}_j\in {\rm I\!R}^3; \alpha_j \in {\mathcal C};  j=1,2,\ldots,N \},
\end{equation}
where $N$ is the number of atoms, $\mathbf{r}_j$ and $\alpha_j$ are, respectively, coordinates and element type of the $j$th atom. Similarly to the discussion in the differential geometry representation section, we only consider non-covalent interactions represented by correlation kernels
\begin{multline}\label{graphCollInter}
{\mathcal E_{kk'}}= \{\Phi(||\mathbf{r}_i-\mathbf{r}_j||; \eta_{kk'})| \alpha_i  = {\mathcal C}_{k}, \alpha_j  = {\mathcal C}_{k'};  i,j = 1,2,\ldots,N;
||\mathbf{r}_i-\mathbf{r}_j||> r_i+r_j +\sigma \},
\end{multline}
all the notations in Eq. \eqref{graphCollInter} are adopted from Sec. \ref{sec:DG}. {  In which,  $\Phi$ refers to  the edge weight which represents the potential interaction between two nodes forming that edge.} We now form weighted colored subgraphs $G({\mathcal V}, {\mathcal E_{kk'}})$ to describe pairwise interactions in a given molecule. To unify the geometric graph-based descriptors for a diversity dataset, we construct multiscale weighted colored subgraph rigidity between $k$th  element type  ${\mathcal C}_{k}$ and  $k'$th element type ${\mathcal C}_{k'}$ {  via a graph centrality type of scheme}
\begin{equation}\label{ESRI3}
 {\rm RI}^G(\eta_{kk'}) =\sum_i \mu^G_i(\eta_{kk'})=\sum_
 {
 {\substack{i \\ \alpha_i  = {\mathcal C}_{k}}}
 }
 \sum_
 {
 {\substack{j \\ \alpha_j  = {\mathcal C}_{k'} \\ ||\mathbf{r}_i-\mathbf{r}_j||> r_i+r_j +\sigma} }
 }
 \Phi(||\mathbf{r}_i-\mathbf{r}_j||;\eta_{kk'}).
\end{equation}
The proposed subgraph rigidity index ${\rm RI}^G(\eta_{kk'})$ in Eq. \eqref{ESRI3} is the aggregation of the collective subgraph centrality $\mu^G_i(\eta_{kk'})$ which used in our previous B-factor prediction model \cite{DBramer:2018a}. That formulation represents a coarse-grained description at the element-level capturing important physical and biology information in a molecule or biomolecule such as van der Waals interactions, hydrogen bonds, electrostatics, etc. {  This description is scalable, i.e., independent of the size of an individual protein-ligand complex.}  In fact, when describing protein-ligand interactions, the labeled subgraph $G({\mathcal V}, {\mathcal E_{kk'}})$ gives rise to a bipartite graph with its edges connecting protein atoms to ligand atoms. {  The positive and negative eigenvalues of  the adjacency matrix of a bipartite graph   are reflective, which enables us to select only positive or negative eigenvalues in machine learning.}
Moreover, Eq. \eqref{ESRI3} generalized our previous binding affinity prediction model \cite{DDNguyen:2017d} and was utilized for the D3R Grand Challenge 3 \cite{nguyen2019mathematical}.

\subsection{Multiscale weighted colored algebraic subgraphs }

\begin{figure}[!ht]
    \centering
    \includegraphics[width=1.0\textwidth]{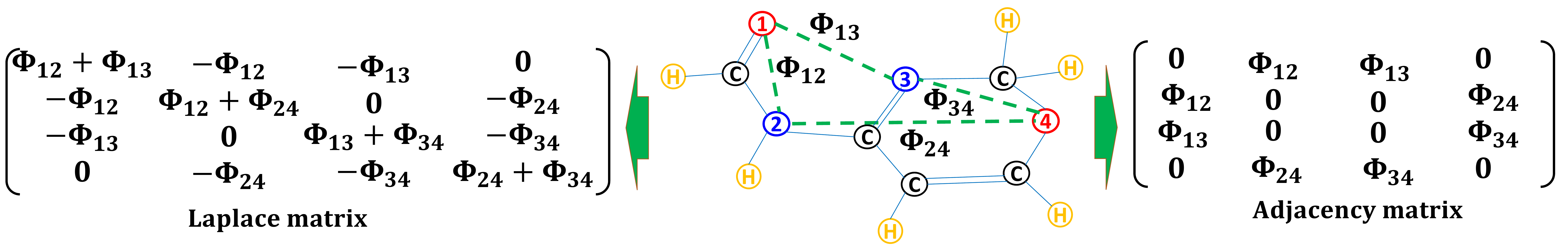}
    \caption{IIlustration of weight colored subgraphs $G_{\rm NO}$ including its Laplacian matrix (Left), and adjacency matrix (Right) deduced from molecule graph (C$_5$H$_6$N$_2$O$_2$) (Middle). Atoms 1 and 4 are oxygen, while atoms 2 and 3 are nitrogen. Graph edges, $\Phi_{ij}$, are in the green-dashed lines representing the noncovalent bonds. In addition, one can get 9 other nontrivial subgraph for this molecule, namely $G_{\rm CC}, G_{\rm CN}, G_{\rm CO}, G_{\rm CH}, G_{\rm NN}, G_{\rm NH}, G_{\rm OO}, G_{\rm OH}, \text{~and~} G_{\rm HH}$.
    }
    \label{fig:AlgebraicGraph}
\end{figure}
Still based on  multiscale weighted colored subgraphs {  as defined in Section \ref{sec:geometric_subgraph}}, we have recently developed a novel algebraic graph approach or spectral graph formulation to describe molecules, biomolecules and their interactions at atomic levels \cite{nguyen2019dg}. We here utilize the Laplacian matrix and adjacency matrix to represent the interactions between nodes in a given subgraph.

Based on a weighted colored subgraph $G({\mathcal V}, {\mathcal E_{kk'}})$, we define the weighted colored Laplacian matrix $L_{ij}(\eta_{kk'})$ as the following

\begin{equation} \label{Laplacianmatrix}
L_{ij}(\eta_{kk'}) = \left\{ \begin{array}{ll}
     - \Phi(||\mathbf{r}_i - \mathbf{r}_j||; \eta_{kk'}) &
     \begin{aligned}
        &\text{if}~  i\neq j, \alpha_i  = {\mathcal C}_{k}, \alpha_j  = {\mathcal C}_{k'}\\
        &\text{and}~  ||\mathbf{r}_i-\mathbf{r}_j||> r_i+r_j +\sigma ;
     \end{aligned}
        \\
     -\sum_j L_{ij}  &  {\rm if}~~ i=j .
        \end{array} \right.
\end{equation}
Due to the symmetric, diagonally dominant and positive-semidefinite, all eigenvalues of the Laplacian matrix $L_{ij}(\eta_{kk'})$ are nonnegative. Moreover, the smallest eigenvalues are zero. It is worth noting that the number of zero eigenvalues can equally referred to the zero-dimensional topological invariant which implies the number of the connected components in the graph. If a graph is connected, there exists one non-zero eigenvalue. Moreover, the smallest non-zero ones is called as Fiedler value representing algebraic connectivity. It is interesting to see that one can reconstruct the geometric graph rigidity via the following formulation
$${\rm RI}^G(\eta_{kk'})={\rm Tr} L(\eta_{kk'}),$$

In addition, we can form the adjacency matrix $A_{ij}$ for the aforementioned subgraph $G({\mathcal V}, {\mathcal E_{kk'}})$ by
\begin{equation} \label{adjacencymatrix}
A_{ij}(\eta_{kk'}) = \left\{ \begin{array}{ll}
     \Phi(||\mathbf{r}_i - \mathbf{r}_j||; \eta_{kk'}) &
    \begin{aligned}
        &\text{if}~  i\neq j, \alpha_i  = {\mathcal C}_{k}, \alpha_j  = {\mathcal C}_{k'}\\
        &\text{and}~  ||\mathbf{r}_i-\mathbf{r}_j||> r_i+r_j +\sigma ;
    \end{aligned}
        \\
   0 &  {\rm if}~~ i=j .
        \end{array} \right.
\end{equation}

  Clearly,   adjacency matrix $A(\eta_{kk'})$  is a  symmetric non-negative matrix. As a result, its spectrum is real. The Laplacian and adjacency matrices for subgraph including only oxygen and nitrogen atoms in  molecule C$_5$H$_6$N$_2$O$_2$ are depicted in Fig. \ref{fig:AlgebraicGraph}. {  Note that for different molecules, one can expect to have different graph structures.}

In general, the element-level information decoded from the Laplacian matrix and the adjacency matrix is quite similar despite of the different behaviors among their eigenvalues and eigenvectors. Specifically, the correlation between the adjacency matrix and the Laplacian matrix can be found in the Perron-Frobenius theorem via the following inequalities
\begin{equation}
 \min_i\sum_j A_{ij} \leq \rho(A)\leq \max_{i} \sum_j A_{ij}.
\end{equation}
In other words, one can state that the spectral radius $\rho(A)$ of the adjacency matrix $A$ is bounded by diagonal element interval of the corresponding Laplacian matrix $L$.

In the algebraic approach, we are interested in describing the interactions between elements in the subgraph by the eigenvalues of its matrix.  Thus, we design the  weighted colored Laplacian matrix based descriptor at the element-level by
\begin{equation}
{\rm RI}^L(\eta_{kk'}) =\sum_i \mu^L_i(\eta_{kk'}),
\end{equation}
and the weighted colored adjacency matrix based descriptor is proposed in a similar manner. Note that GNM \cite{Bahar:1997} is a special case of the proposed Laplacian matrix $\mu^L_i(\eta_{kk'})$. Thus, one can utilize its spectrum $\mu^L_i(\eta_{kk'})$ for the protein B-factor prediction. To enrich the algebraic graph-based description information, we consider the statistics of the eigenvalues such as sum, mean, maximum, minimum and standard deviation.

\subsection{Algebraic topology-based molecular signature}
By employing powerful topological analysis, one can construct sophisticated topological spaces to capture the key interactions at the element level of an interested molecule or biomolecule. These physical and chemical information are encoded in different dimensional space under the topological invariant features, so-called Betti numbers. Upon the topological information, the rich and systematic descriptions are formulated and integrated with advanced machine learning framework.

\subsubsection{Persistent homology}

In the geometric point of view, the collection of points, edges, triangles, and higher-dimension representations form topological spaces. The general form of a triangle or a tetrahedron is called a simplex. Mathematically, a set of $(k+1)$ affinely independent points in $\mathbb{R}^n$ with $n\geq k$ gives rise to a simplex. To further characterize the topological spaces, face is introduced as a convex hull of a subset of points defining a simplex. In addition, a finite collection of  simplices defines a simplicial complex $X$ provided that two requirements are met. First, the faces of any simplex in $X$ are also in $X$. Second, the intersection of two simplices $\sigma_1$ and $\sigma_2$ in $X$ are either empty or a face of both $\sigma_1$ and $\sigma_2$. In a given simplicial complex $X$, a $k$-chain $c$ is a formal sum of all the $k$-simplices in $X$ {  which is defined as $c=\sum_i a_i\sigma_i$. Here, $a_i$ is an integer coefficient chosen in a finite field $\mathbb{Z}_p$ with a prime $p$}. With the additional operator on the coefficients of in the $k$-chain, one can form a group of $k$-chain denoted $\mathcal{C}_k(X)$. The boundary operator on simplices is defined as

\begin{equation}
\partial_k(\sigma) = \sum\limits_{i=0}^k(-1)^i[v_0,\cdots,\hat{v}_i,\cdots,v_k],
\end{equation}
where $v_0,\cdots,v_k$ are vertices of the $k$-simplex $\sigma$ and $[v_0,\cdots,\hat{v}_i,\cdots,v_k]$ means the codim-$1$ face of $\sigma$ be omitting the vertex $v_i$. The boundary operator $\partial_k(\sigma)$ is homeomorphisms going from $\mathcal{C}_{k}(X)$ to $\mathcal{C}_{k-1}(X)$ with an important property $\partial_k\circ\partial_{k+1}=0$. Therefore, one can form the following  chain complex

\begin{equation}
    \cdots
    \xrightarrow{\mathmakebox[.5cm]{\partial_{i+1}}} \mathcal{C}_{i}(X)
    \xrightarrow{\mathmakebox[.5cm]{\partial_{i}}} \mathcal{C}_{i-1}(X)
    \xrightarrow{\mathmakebox[.5cm]{\partial_{i-1}}}
    \cdots
    \xrightarrow{\mathmakebox[.5cm]{\partial_{2}}} \mathcal{C}_{1}(X)
    \xrightarrow{\mathmakebox[.5cm]{\partial_{1}}} \mathcal{C}_{0}(X)
    \xrightarrow{\mathmakebox[.5cm]{\partial_{0}}} 0.
\end{equation}

\begin{figure}[!ht]
    \centering
    \includegraphics[width=0.8\textwidth]{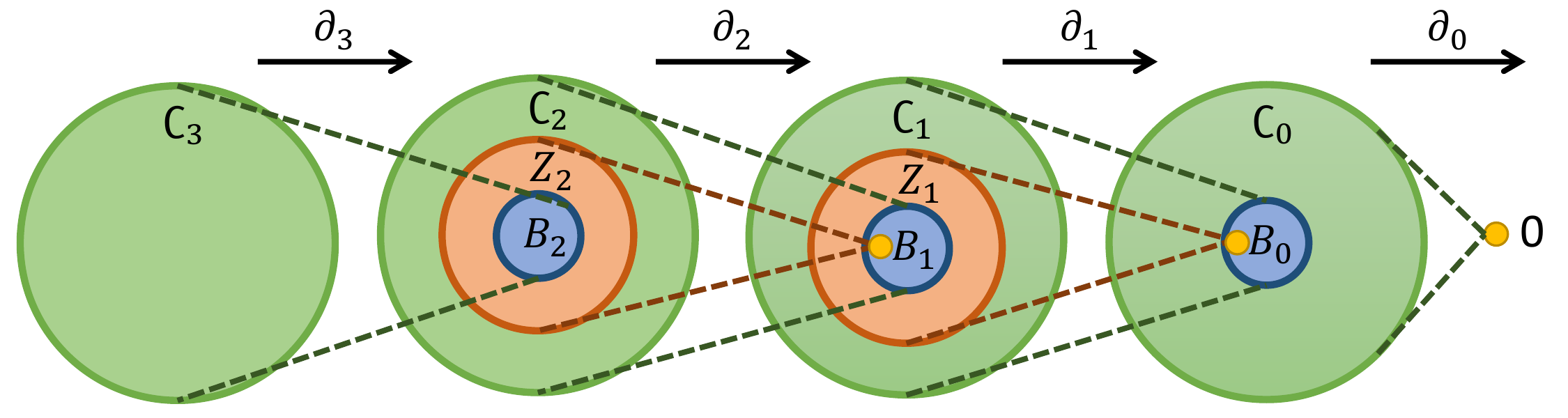}
    \caption{{ Illustration of boundary operators, chain, cycle, and boundary groups in $\mathbb{R}^3.$ Yellow circles are empty sets.}}
    \label{fig:BoundaryOperator}
\end{figure}

In algebraic topology, homology is used to distinguish two shapes by detecting their holes. To define $k$th homology group, we consider the image of the boundary operator $\partial_{k+1}$ denoted $\mathcal{B}_k(X)=\mathrm{Im}(\partial_{k+1})$ and the kernel of  $\partial_{k}$ denoted $\mathcal{Z}_k(X)=\mathrm{Ker}(\partial_k)$ { which are all illustrated in Fig. \ref{fig:BoundaryOperator}.} Then, the quotient group between the aforementioned kernel and image gives rise to the $k$th homology group

\begin{equation}
\mathcal{H}_k(X)=\mathcal{Z}_k(X)/\mathcal{B}_k(X).
\end{equation}

The described above homology group is applied for a fixed topological space. To accommodate the objects related to multiscale, we can construct a sequence of subspaces of topological space. Such sequence is called a filtration $\emptyset = X_0\subseteq X_1\subseteq \cdots\subseteq X_{m-1}\subseteq X_m=X$ which naturally induces a series of homology groups of different dimensions connected by  homomorphisms
\begin{equation}
I_k^{t,s}: \mathcal{H}_k(X_t) \rightarrow \mathcal{H}_k(X_s), \text{~with~} 0\leq t\leq s\leq m.
\end{equation}
The images of these { homomorphisms are} called $k$th persistent homology groups, and ranks of these groups define $k$th persistent Betti numbers {  which are used to recognize topological spaces via number of $k$-dimensional holes. In the physical interpretation, Betti-0 counts the number of independent components, Betti-1 illustrates number of rings, and Betti-2 encodes the cavities. }

\subsubsection{Topological description of molecular systems}

We carry out  persistent homology on labels subgraph $G({\mathcal V}, {\mathcal E_{kk'}})$ defined in the previous sections  to describe molecular properties. The resulting topological formulation is called element specific persistent homology  \cite{ZXCang:2017a,ZXCang:2017b}.

There are two common types of filtration, namely Vietoris-Rips complex and alpha complex \cite{Alpha}. The Vietoris-Rips complex, a distance-based filtration, is used to directly address the protein-ligand interactions. For a set of atoms in  subgraph $G({\mathcal V}, {\mathcal E_{kk'}})$, the subcomplex associated to $\epsilon$ is defined as

\begin{equation}
X_{\rm Rips}(\epsilon) = \{\sigma\in X| \sigma=[v_0,\cdots,v_k],\, d(v_i,v_j) \leq 2\epsilon\,\text{~for~}\,0\leq i,j\leq k \},
\end{equation}
where $X$ is the collection of all possible simplices, $d$ is the distance between two atoms.  To capture a complex protein geometry, one can utilize alpha complex. The alpha filtration is built upon the non-empty intersection between a $k$-simplex and a $(k+1)$ Voronoi cells. In general, in the alpha filtration, the subcomplex associated to $\epsilon$ is defined as
\begin{equation}
X_{\rm alpha}(\epsilon)=\{\sigma\in X| \sigma=[v_0,\cdots,v_k],\, \cap_i\left(V(v_i)\cap B_\epsilon(v_i)\right)\neq \emptyset \},
\end{equation}
where $V(v_i)$ is the Voronoi cell of $v_i$ and $B_\epsilon(v_i)$ is an $\epsilon$ ball centered at $v_i$. { For the details of building an alpha filtration, we refer the interested readers to our published work \cite{KLXia:2014c}.}

Similarly to multiscale weight colored subgraphs in algebraic graph theory approaches, the element specific persistent homology has been shown to capture crucial physical interactions by tweaking the distance functions used in the filtration \cite{ZXCang:2017a,ZXCang:2017b}. Indeed, the hydrophobic effects can be described by considering the persistent homology computation on the collection of all carbon atoms. To describe the hydrophilic behavior of the molecular system, the element specific persistent homology is carried out only for nitrogen and oxygen atoms. In addition, an appropriate distance function selection can characterize the covalent bonds and non-covalent interactions in small molecules \cite{ZXCang:2018a}.

There are several ways to incorporate barcodes generated by persistent homology into machine learning models. One can use the Wasserstein  metric to measure the similarities between two molecules' barcodes. As a result, the distance-based machine learning approaches such as nearest neighbors and kernel methods can be exploited \cite{ZXCang:2018a}.  To make use  advanced machine learning algorithms such as the ensemble of trees and deep neural networks, we vectorize persistent homology barcodes by discretizing them into bins and taking into account of the persistence, birth and death incidents in each bin. Furthermore, the statistics of element-specific persistent homology barcodes are included in fixed length features \cite{ZXCang:2018a}. In the convolutional neural networks, such featurization of barcodes is represented in 1-dimensional and 2-dimensional like images \cite{ZXCang:2017c,ZXCang:2018a}.

\subsection{MathDL energy prediction models}
\begin{figure}[!ht]
    \centering
    \includegraphics[width=0.8\textwidth]{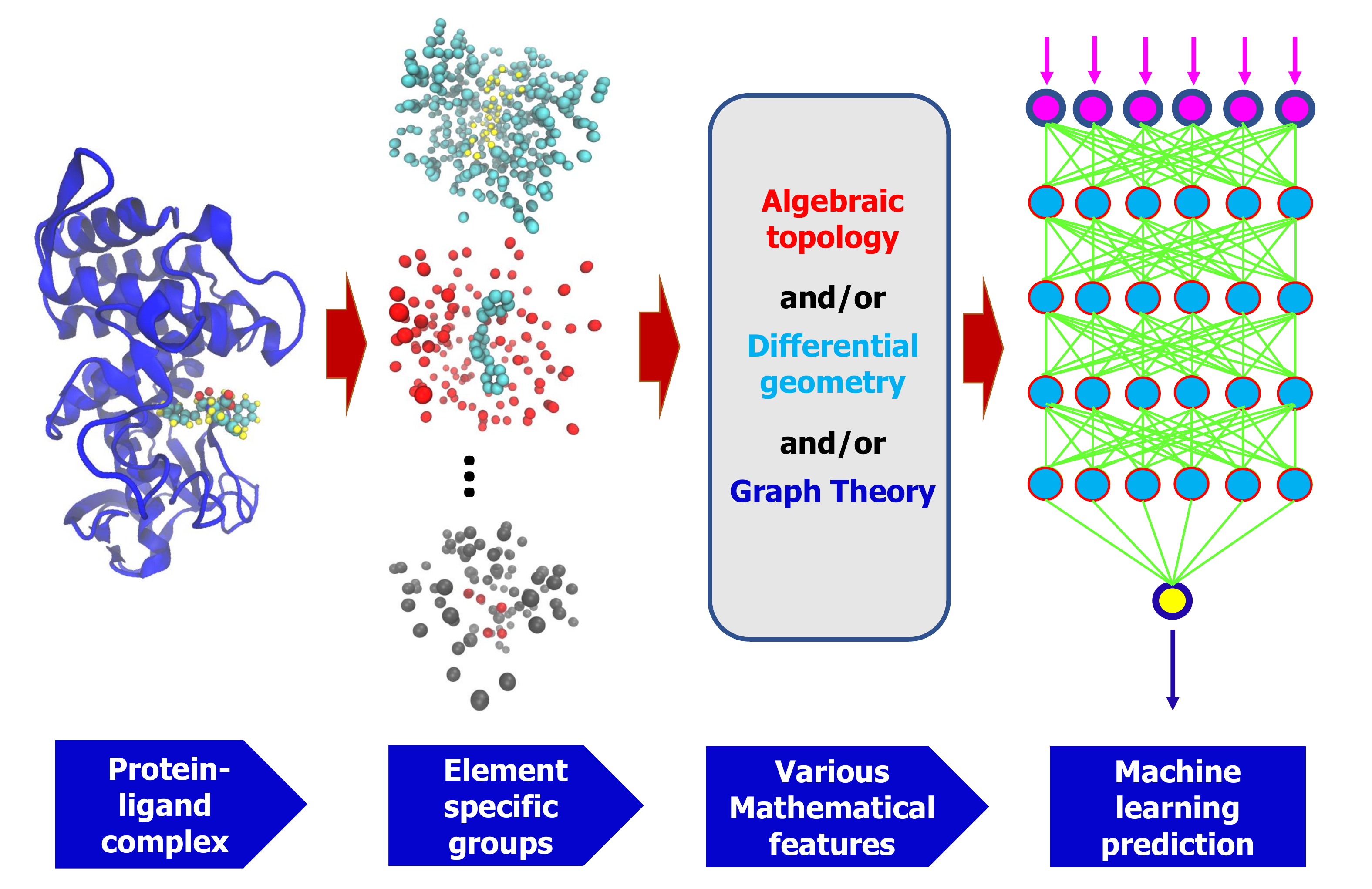}
    \caption{A framework of MathDL energy prediction model which integrates advanced mathematical representations with sophisticated CNN architectures}
    \label{fig:MathDeep}
\end{figure}
We integrate the mathematical features with deep learning networks to form a powerful predictive model. The convolutional neural network (CNN) is a well-known algorithm with much success in image recognition and computer vision analysis. Essentially, CNN is a regularized version of the artificial neural network consisting of many convolutional layers, followed by several fully connected layers. To enhance the learning process, dropout techniques have been exploited in network layers \cite{srivastava2014dropout}. The neural networks we use are classified as the feed-forward network where all the information in the current layer is linearly combined and then nonlinearized via an activation function before sending out to the next layer. The predictive power of the CNN models relies on the characterization of the local interactions in the spatial dimension under the discrete convolution operator. The choice of features inputs in the CNN networks gives rise to variants of binding energy predictive models. Fig. \ref{fig:MathDeep} depicts MathDL energy prediction models {  and their network architectures are described in Fig. S1 in the Supporting Information}. In the D3R GC4, we utilized two different models. In the first approach, the combination of algebraic topology and differential geometry features were employed in the network, we named this model BP1. In the second approach, algebraic topology, differential geometry, and algebraic graph representations were mixed to lead to another binding energy prediction model named BP2. {  The details of feature generation procedure of the algebraic topology, differential geometry, and algebraic graph models can be found in our earlier work \cite{ZXCang:2018a,nguyen2019dg,nguyen2019agl}}.

\subsection{MathDeep  docking models}
\begin{figure}[!ht]
    \centering
		\includegraphics[width=0.5\textwidth]{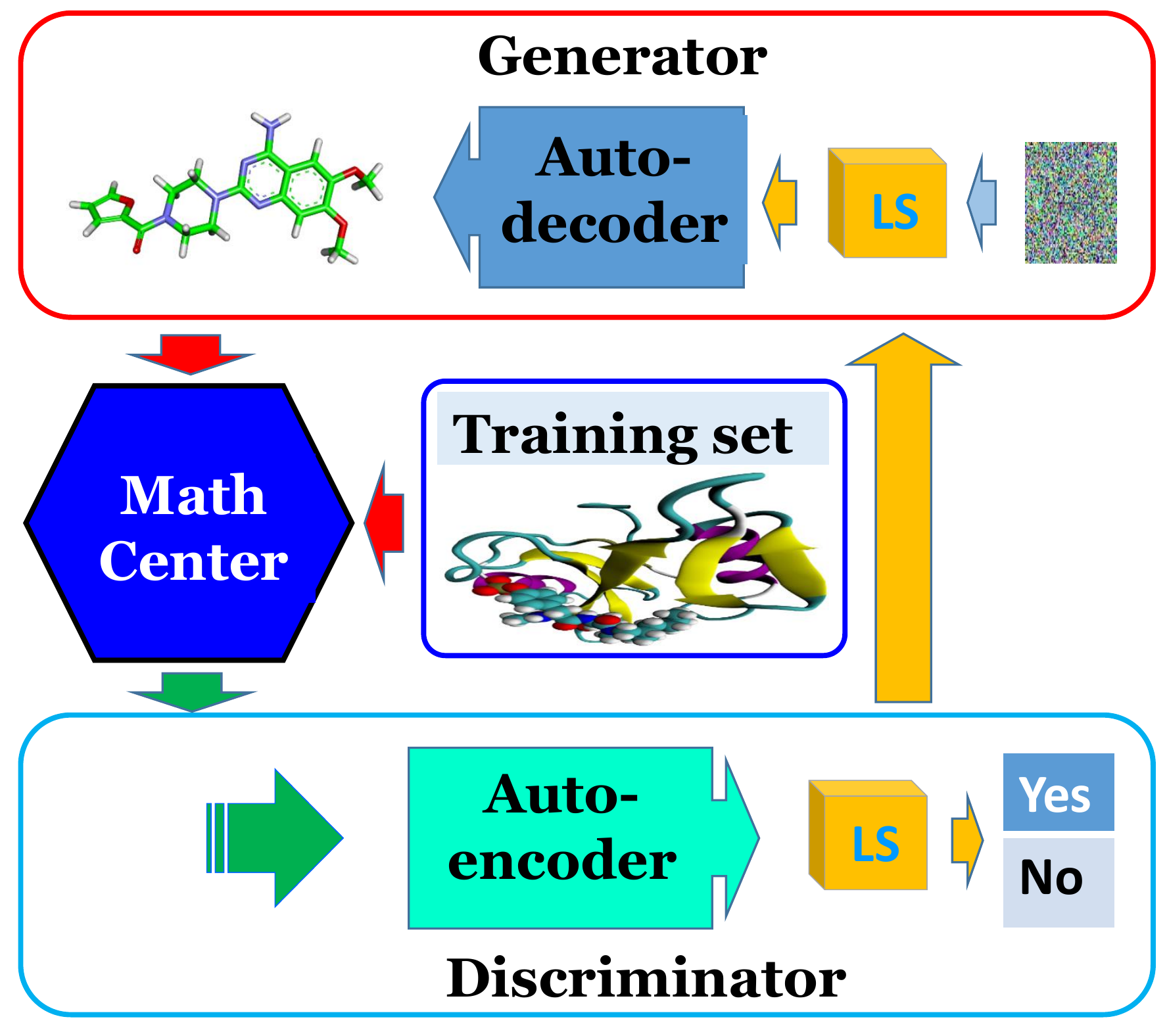}
    \caption{ Illustration of our docking approach using mathematical representations integrated with GAN architectures.  The generator contains an autodecoder, a latent space (LS), and a noise source. The discriminator consists of an autoencoder and latent space. The Math center encodes 3D structures into low-dimensional mathematical representations using algebraic topology, differential geometry, and/or graph theory. }
    \label{fig:MathDockGAN}
\end{figure}
We here present an innovative pose generation scheme, denoted MGAN, using advanced mathematical representation pre-conditioned generative adversarial networks (GAN). GAN is a kind of deep learning model consisting of a generator $G$ { to learn} the data distribution, and a discriminator $D$ to discriminate training set structural information from that of  the generator $G$  \cite{goodfellow2014generative}. The $G$ model is iteratively improved from the $D$ feedback until the $D$ cannot tell the difference between training set structural information and $D$ set one.
To improve the GAN performance  and avoid vanishing gradient and mode collapse, we employ Wasserstein GAN (WGAN) \cite{arjovsky2017wasserstein} in our model. To further enhance the quality of the generated structures, we take advantage of the conditional GAN { technique} \cite{mirza2014conditional}. The deep learning (DL) models $G$ and $D$ are partially adapted from our binding energy prediction networks which are  fed with data encoded in intrinsically low-dimensional manifolds with differential geometry, algebraic topology and graph theory.    Fig. \ref{fig:MathDockGAN} depicts the MGAN's framework. { Network architectures of autodecoder and autoencoder are illustrated in  Figs. S2 and S3, respectively.} By varying combinations of different mathematics, we end up with several docking models. Specifically, If DL networks $G$ and $D$ only exploit algebraic topology, we name this docking model DM1. Similarly, we attain DM2 and DM3  when GAN model includes only algebraic graph and differential geometry based representations, respectively. Finally, DM4 is constructed with the assistance of  algebraic topology, algebraic graph, and differential geometry.
{  We employed the PDBbind v2018 dataset to train MathDL and MGAN models.}
{  The optimal hyperparameters of the MathDL model were selected by experience and finalized by hyperopt python package (\url{http://github.com/hyperopt/hyperopt}). The MGAN model was trained based on the setting of Wasserstein GAN network discussed in this work \cite{arjovsky2017wasserstein}. Furthermore, }
{ to enhance the pose generation quality, we carry out the transfer learning to further optimize the MGAN model with the protein family-specific structures.}

\section{Results and discussion}

In this section, we present MathDL results and discuss our performances in the latest Grand Challenge named GC4.

\subsection{Pose prediction results and discussion }
\begin{figure}[!ht]
    \centering
    \includegraphics[width=\textwidth]{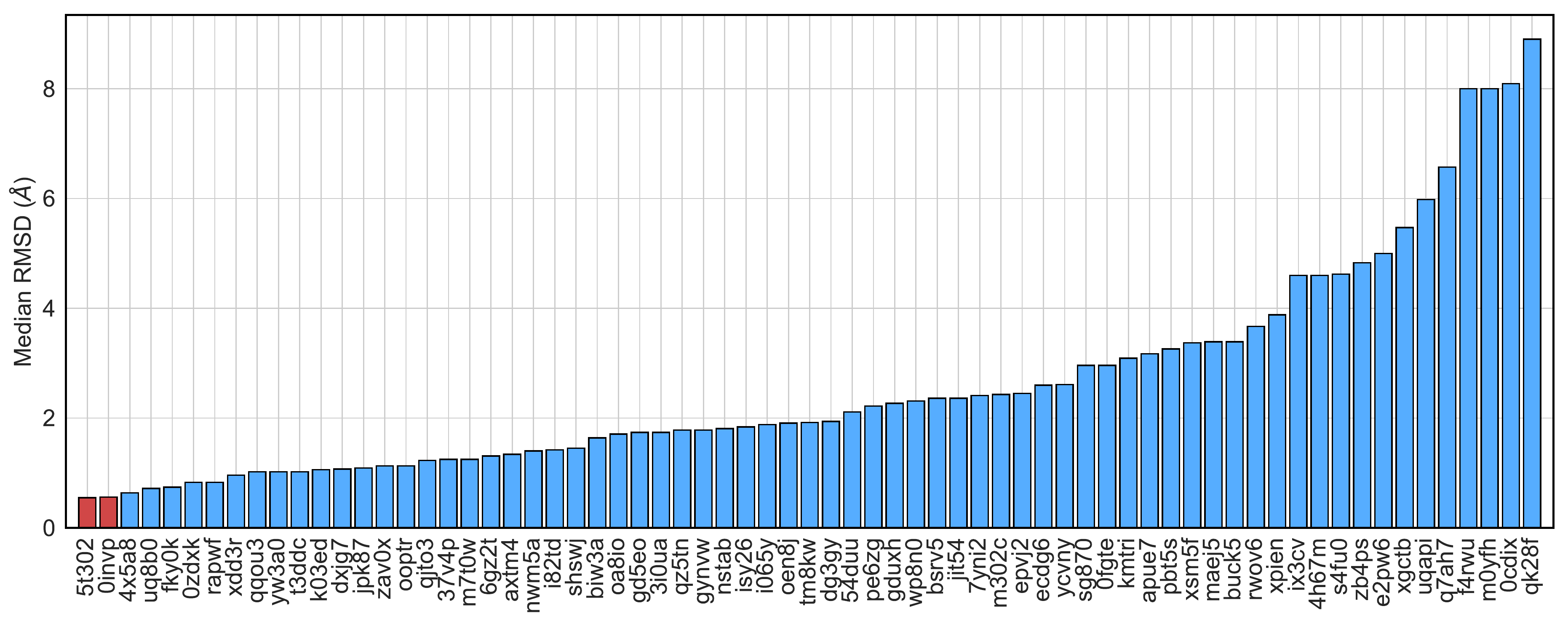}
    \caption{Performance comparison of different submissions on pose prediction challenge of Stage 1a {  for the BACE dataset} in term of median RMSD. { Our submissions are highlighted in the red color, in which the best one is 5t302 with median RMSD = 0.55 \AA~.}
    }
    \label{fig:stage1a_BACE_pose}
\end{figure}

\begin{figure}[!ht]
    \centering
    \includegraphics[width=0.7\textwidth]{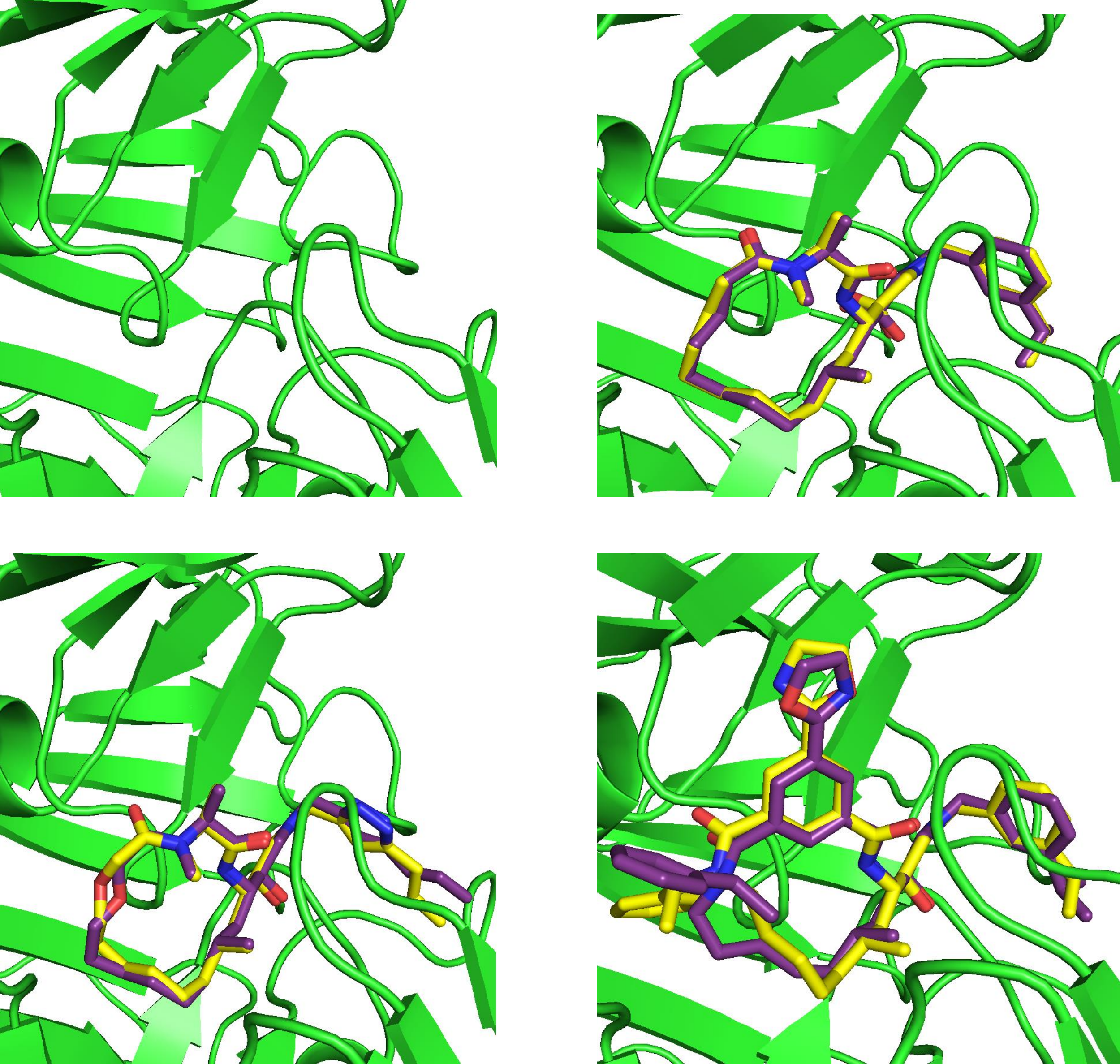}
    \caption{Illustration of pose predictions by our MathGAN docking model with receipt ID 0invp. The top-left corner is original binding pocket of the BACE receptor. The top-right corner is our best pose prediction accuracy obtained when predicting BACE03's pose with RMSD = 0.23 \AA~. The bottom-left corner is our middle performance when predicting BACE05's pose with RMSD = 0.53 \AA~. The bottom-right is our worst performance when predicting BACE07's pose with RMSD = 2.63 \AA~. The experiment structures are in yellow while the predicted structures are in purple.
    }
    \label{fig:BACE_binding_poses}
\end{figure}
We have participated in the docking challenge task since D3R GC2. Before the current challenge, i.e., GC4, our docking results in term of RMSE were not competitive in comparison to those of other participants. Specifically, our mean RMSD values are 6.03 \AA~ and 3.78 \AA~ for GC2 and GC3, respectively. These results reflect an improvement in our docking approaches but their accuracy is still behind the top submissions in GC3. Instead of depending on the docking programs such as Autodock Vina \cite{Trott:2010AutoDock} and GLIDE\cite{Friesner:2004Schrodinger} as we did in the previous challenges, our GC4 docking schemes were driven by advanced mathematical representations and sophisticated deep learning architectures. Consequently, we achieved remarkable performances on the pose prediction tasks. The rest of this section is devoted to  result discussions.

Despite having two protein receptors in GC4, all the pose predictions were only for BACE ligands and were organized in two stages, Stage 1a and Stage 1b. In Stage 1a, participants were provided SMILES strings of 20 ligands to be docked, the FASTA sequence of the BACE protein, and the reference protein structure (PDBID: 5ygx, chain A) for the superimposition process.  Stage 1b took place right after the end of Stage 1a. {  Stage 1b provided the experimental protein structures in the complexes with 20 ligands requested for pose predictions, in which the structures of these ligands were removed}. Participants were still asked to predict their poses. Therefore, Stage 1b is often referred to a self-docking challenge. There are two evaluation metrics for the pose prediction tasks, namely median and mean calculated over all RMSD values between the predicted poses and crystal structures.

In Stage 1a, we submitted two results.
Fig. \ref{fig:stage1a_BACE_pose} illustrates the performances of 70 submissions having median RMSD less than 10 \AA~. Our best submission having receipt ID {  5t302} with median RMSD = 0.53 \AA~ and being highlighted in the red color. This docking model was { DM1}. In Stage 1b, we delivered 4 submissions; unfortunately, none of them was ranked the first place in either the median or mean metric. However, our results were very promising. Particularly, our submission based on docking model DM3 with receipt ID itzv6 achieved mean RMSD of 0.73 \AA~ which is at the second place and is a bit less accurate than the top submission with mean RMSD being 0.61 \AA~ (receipt ID 5od5g). It may be  noted that the best result in Stage 1b is not as good as that in Stage 1a. Fig. \ref{fig:BACE_binding_poses} compares the poses predicted by our submission ID 0invp to the corresponding experimental structures at different levels of accuracy.

It is interesting to find out that, the additional information of the co-crystal structures did not help our docking models. For example, our docking approach DM4 with submission ID Oinvp attained median RMSD of 0.53 \AA~ and mean RMSD of 0.8 \AA~, respectively in Stage 1a. However, in Stage 1b, the same model labeled by receipt ID 2ieqo produced median RMSD and mean RMSD as high as 0.55 \AA~ and 0.84 \AA~, respectively. These observations can confirm the robustness of our models and predictive value for the realistic situations in CADD when  little or no co-crystal information is provided.

\subsection{Affinity prediction results and discussion}

\begin{figure}[!ht]
    \centering
    \includegraphics[width=\textwidth]{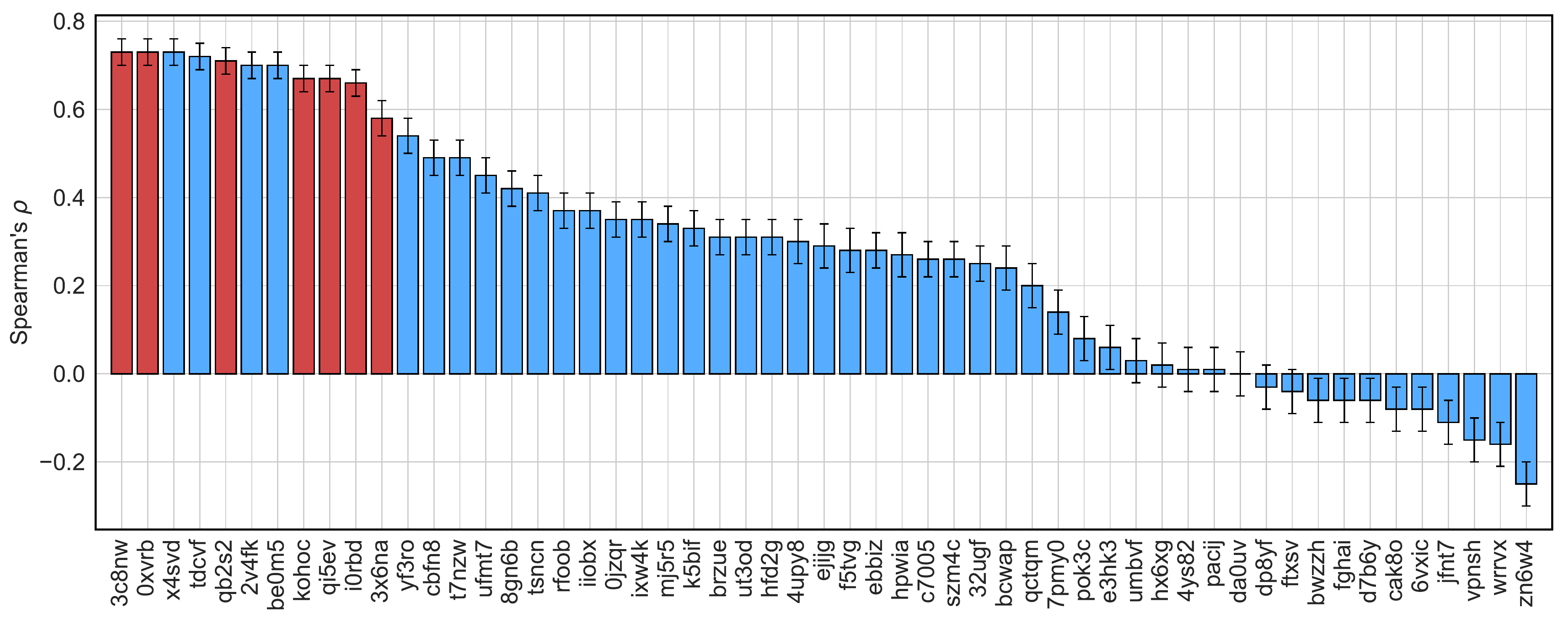}
    \caption{Performance comparison of different submissions on the combined ligand and structure based scoring of CatS dataset in term of Spearman's $\rho$. { Our submissions are highlighted in the red color, in which our top-ranked submissions are 3c8nw and 0xvrb with $\rho$=0.73.}
    }
    \label{fig:stage_affinityranking_combined_CatS}
\end{figure}

\begin{figure}[!ht]
    \centering
    \includegraphics[width=\textwidth]{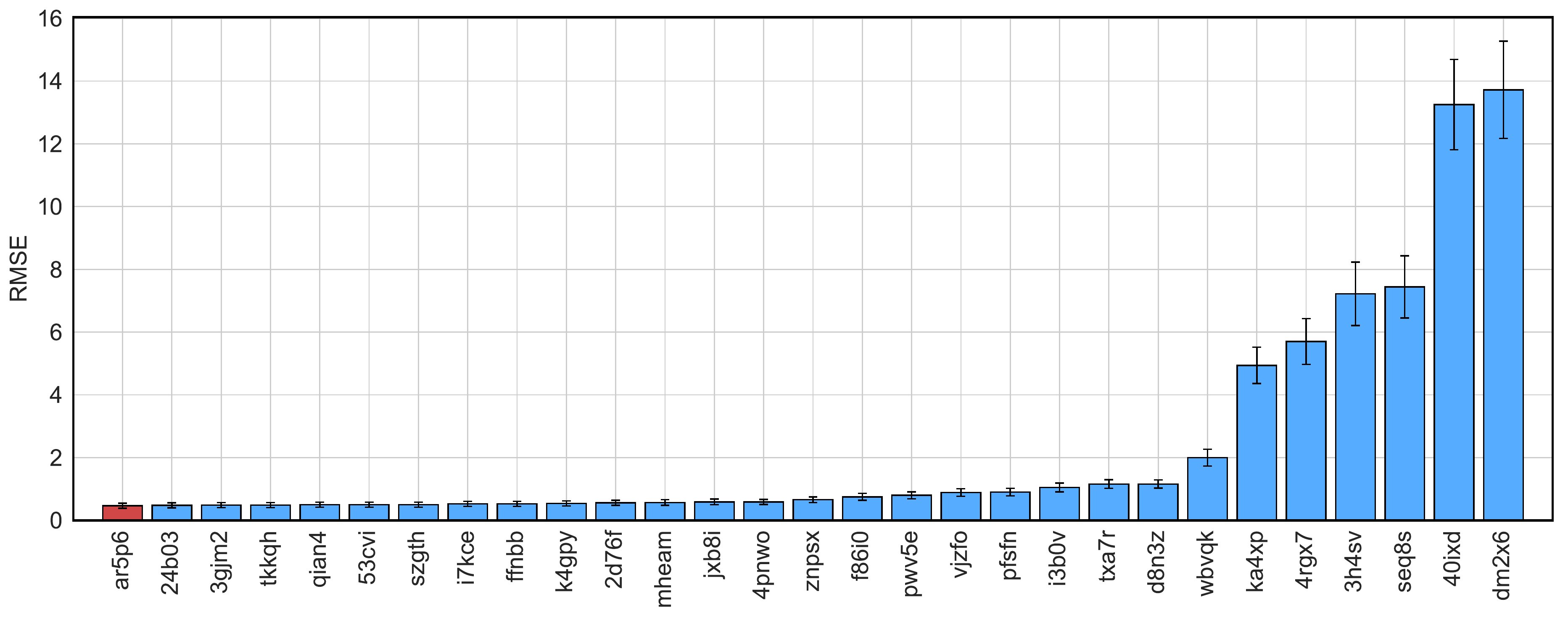}
    \caption{Performance comparison of D3R GC4 participants on free energy set for CatS contest in term of centered RMSE $\text{RMSE}_c$. { Our submissions are highlighted in the red color, in which our top-ranked prediction is ar5p6  with $\text{RMSE}_c$ = 0.47 kcal/mol.}
    }
    \label{fig:stage_FE_CatS}
\end{figure}

\begin{table*}[!ht]
   \small
    \centering
    \caption{
        Overview of all predictive tasks in D3R GC4. Gold medal, silver medal, and bronze medal indicate the ranking of MathDL predictions are first, second, and third, respectively. The numbers $(a/b)$ right beside each medal, say gold medal, implies we have $a$ predictions  were ranked 1st and there was a total of $b$ submissions sharing the first position.
    }
    \scalebox{0.9}{
    \begin{tabular}{llc} \toprule
        {Dataset} & {Contest}  & Results\\ \midrule
        {\bf Pose Prediction} & & \\
        BACE Stage 1A & Pose Prediction & \includegraphics[width=0.03\linewidth]{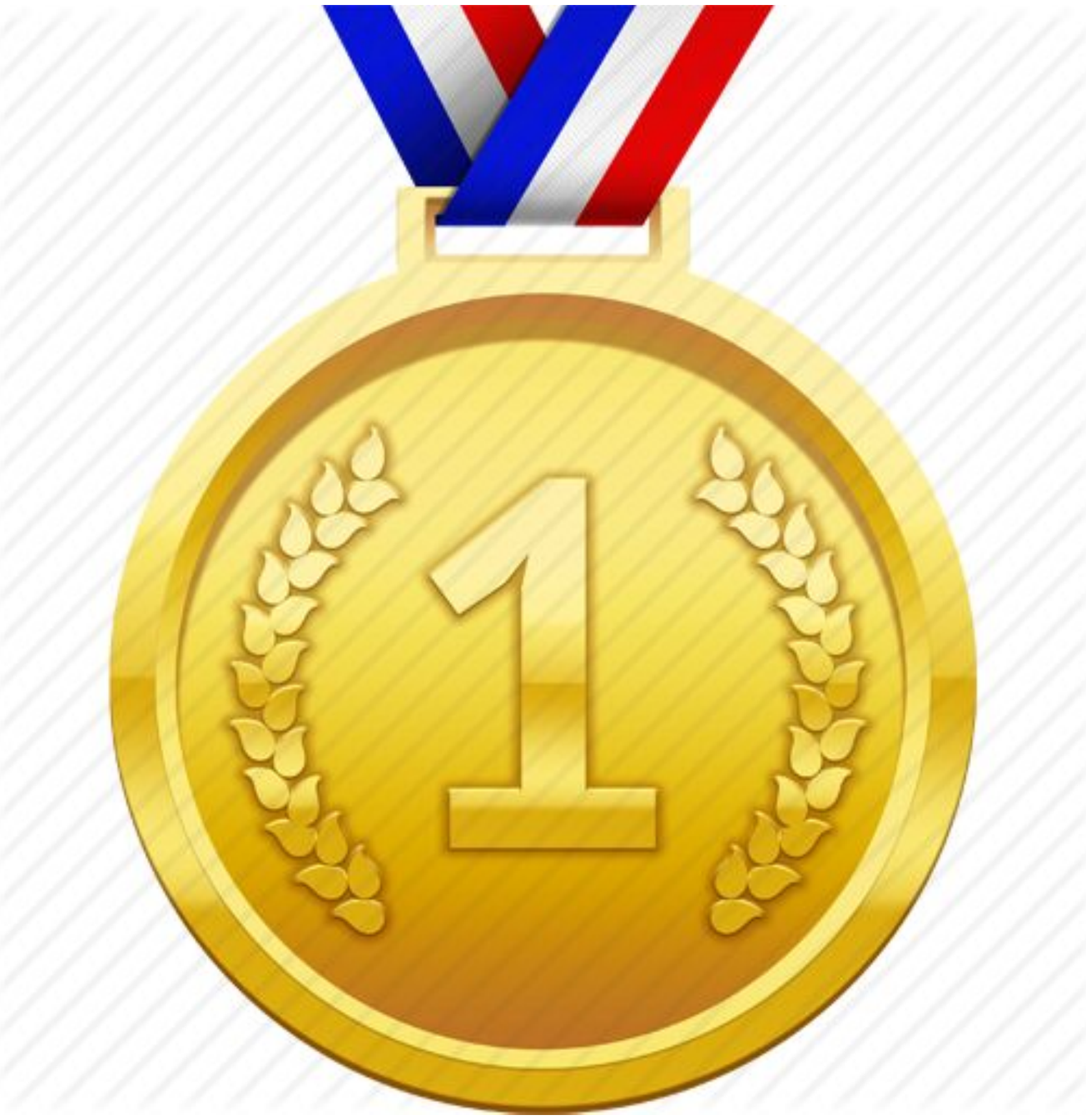} (1/2)$^{\rm  i}$ \includegraphics[width=0.03\linewidth]{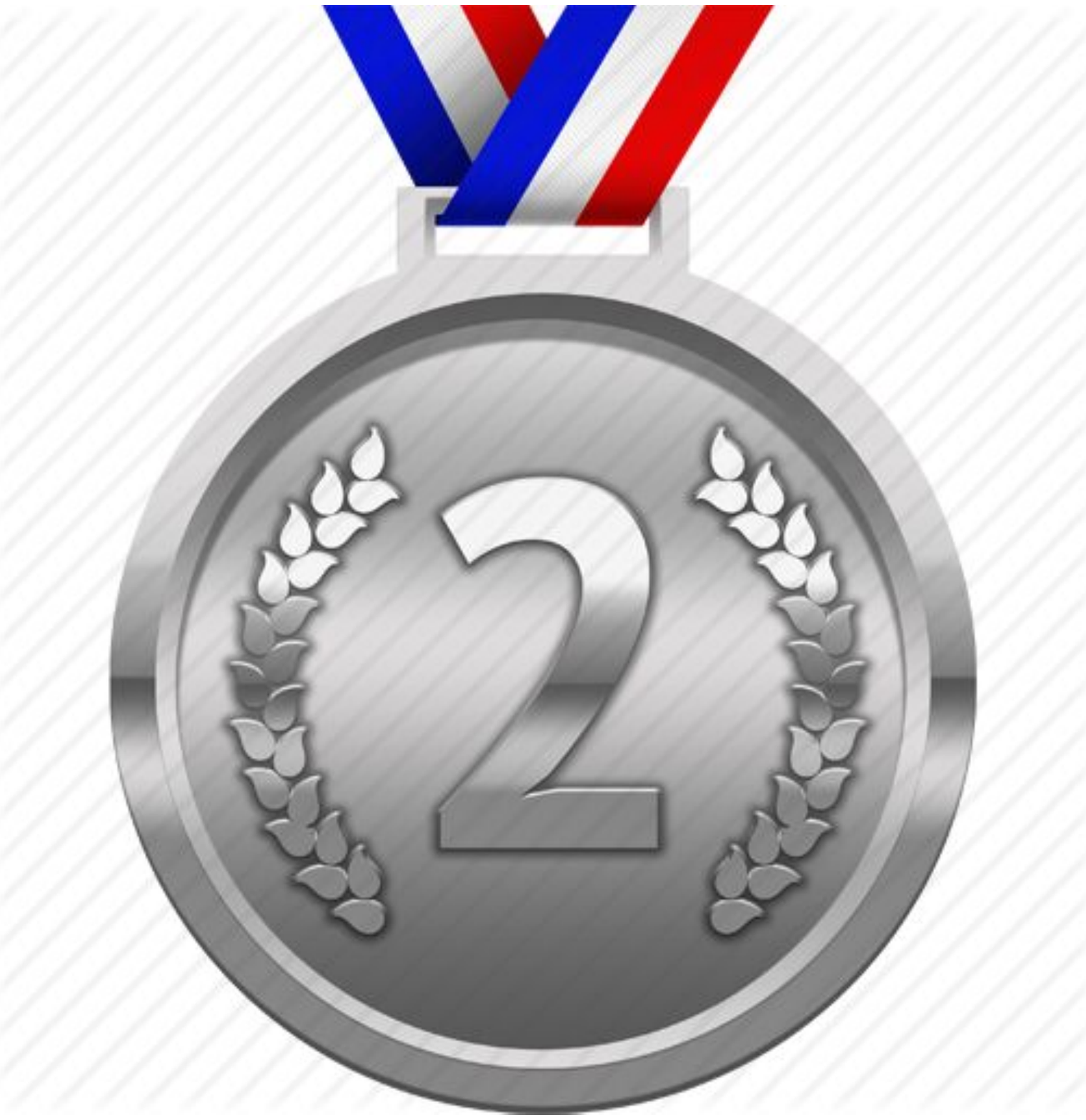} (3/3)$^{\rm  ii}$ \\
        BACE Stage 1B & Pose Prediction & \includegraphics[width=0.03\linewidth]{silver-medal.pdf} (2/2)$^{\rm  iii}$ \includegraphics[width=0.03\linewidth]{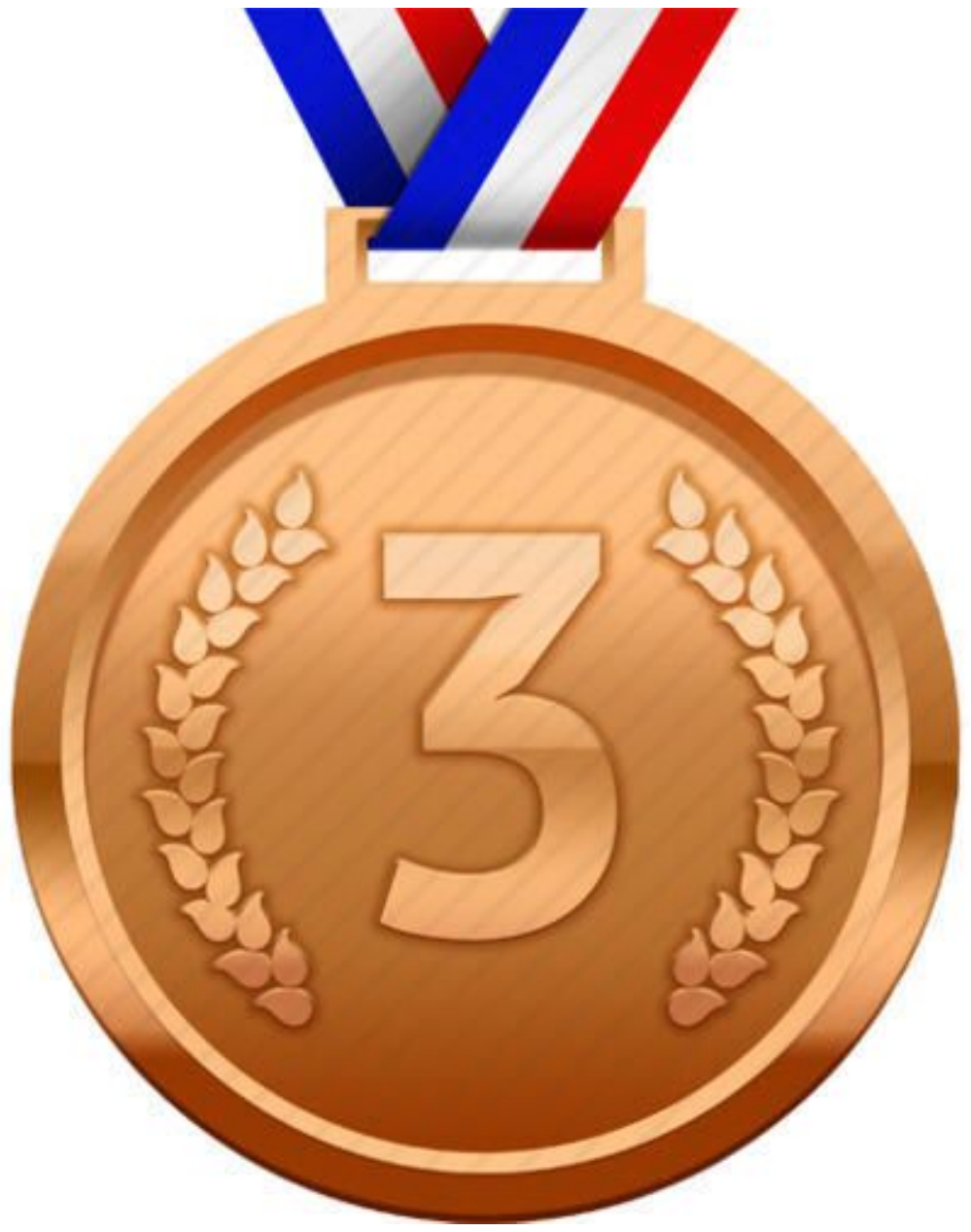} (1/2)$^{\rm  iv}$ \\[1pt] \\
        {\bf Affinity Predictions} & & \\
        Cathepsin Stage { 2} & Combined Ligand and Structure Based Scoring & \includegraphics[width=0.03\linewidth]{gold-medal.pdf} (2/5)$^{\rm  v}$ \includegraphics[width=0.03\linewidth]{silver-medal.pdf} (2/3)$^{\rm  vi}$ \includegraphics[width=0.03\linewidth]{bronze-medal.pdf} (2/4)$^{\rm  vii}$ \\
        Cathepsin Stage { 2} & Ligand Based Scoring & No participation \\
        Cathepsin Stage { 2} & Structure Based Scoring & \includegraphics[width=0.03\linewidth]{gold-medal.pdf} (2/4)$^{\rm  viii}$ \includegraphics[width=0.03\linewidth]{silver-medal.pdf} (3/3)$^{\rm  ix}$ \includegraphics[width=0.03\linewidth]{bronze-medal.pdf} (3/3)$^{\rm  x}$ \\
        Cathepsin Stage { 2} & Free Energy Set & \includegraphics[width=0.03\linewidth]{gold-medal.pdf} (1/7)$^{\rm  xi}$ \includegraphics[width=0.03\linewidth]{silver-medal.pdf} (1/7)$^{\rm  xii}$ \includegraphics[width=0.03\linewidth]{bronze-medal.pdf} ({ 3}/5)$^{\rm  xiii}$ \\
        BACE Stage 1 & Combined Ligand and Structure & No participation \\
        BACE Stage 1 & Ligand Based Scoring & No participation \\
        BACE Stage 1 & Structure Based Scoring & No participation \\
        BACE Stage 1 & Free Energy Set & No participation \\
        BACE Stage 2 & Combined Ligand and Structure &  \\
        BACE Stage 2 & Ligand Based Scoring & No participation \\
        BACE Stage 2 & Structure Based Scoring & \\
        BACE Stage 2 & Free Energy Set &  \includegraphics[width=0.03\linewidth]{silver-medal.pdf} (3/4)$^{\rm  xiv}$ \includegraphics[width=0.03\linewidth]{bronze-medal.pdf} (1/4)$^{\rm  xv}$\\
        \bottomrule
  \end{tabular}
  }
  \scalebox{0.9}{
   
  \begin{tabular}{lclcc}
  Superscript & Submission ID & Evaluation Metric & Docking Protocol & Scoring Protocol \\
    ${\rm i}$   & 5t302 & Median RMSD   & DM1 \\
    ${\rm ii}$  & 5t302 & Mean RMSD     & DM1\\
                & 0invp & Median RMSD   & DM4\\
                & 0invp & Mean RMSD     & DM4 \\
    ${\rm iii}$ & 2ieqo & Median RMSD   & DM4 \\
                & itzv6 & Mean RMSD     & DM3 \\
    ${\rm iv}$  & 4myne & Mean RMSD     & DM1 \\
    ${\rm v}$   & 0xvrb & Spearman's $\rho$     & DM3   & BP2 \\
                & 3c8nw & Spearman's $\rho$     & DM4   & BP2 \\
    ${\rm vi}$  & 0xvrb & Kendall's $\tau$      & DM3   & BP2 \\
                & 3c8nw & Kendall's $\tau$      & DM4   & BP2 \\
    ${\rm vii}$ & qb2s2 & Kendall's $\tau$      & DM1   & BP2 \\
                & qb2s2 & Spearman's $\rho$     & DM1   & BP2 \\
    ${\rm viii}$   & 0xvrb & Spearman's $\rho$     & DM3   & BP2 \\
                   & 3c8nw & Spearman's $\rho$     & DM4   & BP2 \\
    ${\rm ix}$  & 0xvrb & Kendall's $\tau$      & DM3   & BP2 \\
                & 3c8nw & Kendall's $\tau$      & DM4   & BP2 \\
                & qb2s2 & Spearman's $\rho$     & DM1   & BP2 \\
    ${\rm x}$   & qb2s2 & Kendall's $\tau$      & DM1   & BP2 \\
                & qi5ev & Spearman's $\rho$     & DM3   & BP1 \\
                & kohoc & Spearman's $\rho$     & DM2   & BP2 \\
    ${\rm xi}$   & ar5p6 & RMSE$_{\rm c}$       & DM4   & BP2 \\
    ${\rm xii}$  & 24b03 & RMSE$_{\rm c}$       & DM3   & BP2 \\
    ${\rm xiii}$  & 24b03 & Kendall's $\tau$        & DM3   & BP2 \\
                  & 24b03 & Spearman's $\rho$       & DM3   & BP2 \\
                  & 24b03 & Pearson's $r$           & DM3   & BP2 \\
    ${\rm xiv}$  & 8frur & Kendall's $\tau$        & DM1   & BP2 \\
                 & 8frur & Spearman's $\rho$        & DM1  & BP2 \\
                 & 8frur & RMSE$_{\rm c}$           & DM1  & BP2 \\
    ${\rm xv}$  & 8frur & Pearson's $r$            & DM1   & BP2 \\
        \bottomrule
      \end{tabular}
      }
  \label{tab:GC4_results}
  \end{table*}

There were two subchallenges for affinity prediction tasks. Subchallenge 1 regarded BACE ligands while Subchallenge 2 concerned CatS ligands. Both subchallenges were interested in affinity ranking of a diversity datasets and relative binding affinity predictions on the designated free energy set. There were two stages on BACE affinity prediction task, namely Stage 1 and Stage 2, whereas there was only one stage on CatS ligands. { Unfortunately, we did not participate in Stage 1 of the BACE target since the announcement email made us overlook this contest.}

Statistically, there were 154 compounds in the BACE dataset for affinity ranking contest, while there were 34 compounds for the calculation of relative or absolute binding affinities of the same receptor target. In CatS dataset, participants were asked to rank affinities of 459 ligands and predicted the binding energies of a smaller subset with 39 molecules. Moreover, Kendall's $\tau$ and Spearman's $\rho$ were the evaluation metrics for affinity ranking challenges. In the binding free energy predictions, besides the aforementioned metrics, Pearson's $r$ and centered root mean square error (${\text{RMSE}_c}$) were utilized.

Overall, the official results from the D3R organizer have placed us among the top performers on these energy prediction contests. By considering specific evaluation metrics, we were ranked first place in {\it combined ligand and structure based scoring{ \footnote{This subcategory is the common list of ligand based and structure based scoring subcategories}}}, {\it structure based scoring}, and {\it free energy set} subcategories all belonging to the CatS dataset. For illustration, Fig. \ref{fig:stage_affinityranking_combined_CatS} presents the Spearman's $\rho$ performance of different submissions on the CatS affinity ranking contest combining ligand and structure based scoring models. Our best submission are highlighted in the red color with receipt IDs 3c8nw and 0xvrb. Both of them achieved the same Spearman's $\rho$ as high as 0.73 and shared the first place with another group's submission having ID x4svd. In submission ID 3c8nw, we employed docking model DM4 for pose generation and model { BP2} for the affinity prediction. While in submission ID 0xvrb, docking approach was DM3 and binding prediction protocol was { BP2}.
In addition, our best result with ID ar5p6 achieved the lowest $\text{RMSE}_c$ for the free energy prediction of 39 designated CatS molecules. This successful submission utilized docking model DM4 and affinity prediction model { BP2} for the calculations. Fig. \ref{fig:stage_FE_CatS} presents $\text{RMSE}_c$ performance of various groups for the free energy prediction of CatS dataset.
Table \ref{tab:GC4_results} summarizes the performances of our group at all categories in D3R GC4. We only counted the number of our submissions in the top three including ties. ``No participation'' at the results column implies that we did not participate in the corresponding contest. The blank results indicate that our predictions were not ranked within the top three.

{ It is noted} that in the BACE affinity prediction, our results were not in the top three. In fact, our team was behind only to  two teams that collected all the top three places in BACE affinity ranking, which indicates the consistence of our MathDL models in GC4 competitions.

{ Overall, the model BP2 was our best model for binding affinity prediction for both CatS and BACE datasets (see Table S1). The great performance of BP2 was expected since it combines algebraic topology, differential geometry, and graph theory features which help to enrich feature space and cover the most important aspects of physical and biological properties. However, there was a mixed conclusion when finding the best solution for pose prediction. Indeed, models DM3 and DM4 worked well for the CatS dataset, while DM1 was an only good solution for producing high quality poses for the BACE dataset (see Table S1). They helped the predictor BP2 achieved the best rankings among our submitted models. One can argue that DM1 achieved the best pose prediction for BACE ligands in Stage 1A; therefore it was foretasted to help BACE energy prediction tasks. The same behavior was observed for CatS dataset. According to our pre-validation results, DM4 which was our best model for the CatS pose prediction, achieved mean RMSD of 1.8 \AA~ for the CatS pose prediction Stage 1B challenge in GC3. Note that the best submission in that subchallenge accomplished mean RMSD as low as 2.13 \AA. It seems that the pose quality of our pose generation models correlates well to the accuracy of our binding affinity predictors.
}

\section{Conclusion}

The performances of our mathematical deep learning (MathDL) models on D3R GC4 are presented and discussed in this paper. We participated in a variety of D3R GC4 contests including pose predictions, affinity ranking, and absolute free energy predictions. Overall, our submissions were ranked the first in pose prediction in Stage 1a, affinity ranking and free energy predictions for Cathepsin ligands. Unfortunately, we did not get the first place on BACE datasets. Our best submission was only at the second place in free energy set for BACE in Stage 2 contest. In comparison to our previous D3R challenges, i.e., D3R GC2 and D3R GC3, we had two improvements in D3R GC4. The first improvement was the pose prediction. This was the first time we won this contest thanks to our newly developed docking model which integrates {  scalable low-dimensional rotational and translational invariant mathematical representations}, such as differential geometry, algebraic graph, and algebraic topology,  with  well-designed generative adversarial networks.  The second improvement was the affinity ranking for a dataset with diverse chemical properties. In previous challenges, our approaches performed well on free energy predictions but not on affinity ranking. In GC4, we successfully unified our newly established models, i.e., differential geometry and algebraic graph, and our well-known algebraic topology into powerful and robustness convolutional neural network models for binding affinity predictions.

In terms of efficiency, at this point, our MathDL models are quite automated. With  sufficient computer resources, our MathDL models can finish all the GC4 competition tasks in a week or so.

{ It is worth noting }that our models for GC4 was the less competitive performance in BACE affinity ranking and free energy predictions. Additionally, it seems that our docking model  did not upgrade when the co-crystal structures became available.  These issues are under our investigation.

 \section*{Acknowledgments}
This work was supported in part by  NSF Grants DMS-1721024,  DMS-1761320, and IIS1900473 and NIH grant  GM126189. DDN and GWW are also funded by Bristol-Myers Squibb and Pfizer.


\end{document}